\documentclass[aps,prl,twocolumn,superscriptaddress]{revtex4}
\usepackage[dvips]{graphicx}
\def\degC{\kern-.2em\r{}\kern-.3em C }

\begin{document}

%\preprint{}

\title{Valence and Na content dependences of superconductivity in Na$_{x}$CoO$_{2}\cdot y$H$_{2}$O}


\author{Hiroya Sakurai}
\email[Corresponding author: ]{sakurai.hiroya@nims.go.jp}
\affiliation{Superconducting Materials Center, National Institute for Materials Science, Namiki 1-1, Tsukuba, Ibaraki 305-0044, Japan}
\affiliation{International Center for Young Scientists, National Institute for Materials Science, Namiki 1-1, Tsukuba, Ibaraki 305-0044, Japan}

\author{Naohito Tsujii}
\author{Osamu Suzuki}
\author{Hideaki Kitazawa}
\author{Giyuu Kido}
\affiliation{NanoMaterials Laboratory, National Institute for Materials Science, Sengen 1-2-1, Tsukuba, Ibaraki 305-0044, Japan}

\author{Kazunori Takada}
\author{Takayoshi Sasaki}
\affiliation{Advanced Materials Laboratory, National Institute for Materials Science, Namiki 1-1, Tsukuba, Ibaraki 305-0044, Japan}

\author{Eiji Takayama-Muromachi}
\affiliation{Superconducting Materials Center, National Institute for Materials Science, Namiki 1-1, Tsukuba, Ibaraki 305-0044, Japan}

%\thanks{}


\date{\today}

\begin{abstract}
Various samples of sodium cobalt oxyhydrate with relatively large amounts of Na$^{+}$ ions were synthesized by a modified soft-chemical process in which a NaOH aqueous solution was added in the final step of the procedure. From these samples, a superconducting phase diagram was determined for a section of a cobalt valence of $\sim$+3.48, which was compared with a previously obtained one of $\sim$+3.40. The superconductivity was significantly affected by the isovalent exchanger of Na$^{+}$ and H$_{3}$O$^{+}$, rather than by variation of Co valence, suggesting the presence of multiple kinds of Fermi surface. Furthermore, the high-field magnetic susceptibility measurements for one sample up to 30 T indicated an upper critical field much higher than the Pauli limit supporting the validity of the spin-triplet pairing mechanism.
\end{abstract}

%\keywords{}

\maketitle

The sodium cobalt oxyhydrate Na$_{x}$CoO$_{2}\cdot y$H$_{2}$O has been studied intensively since the discovery of a superconducting transition below about 5 K \cite{TakadaNature}. Superconductivity can be induced in a CoO$_{2}$ layer having a triangular lattice of Co and this superconductivity is considered to be of an unconventional type \cite{FujimotoPRL, IshidaJPSJ}. Supporting the existence of this exotic superconductivity, it has recently been discovered that, in this system, a magnetically ordered phase is located just next to the superconducting phase \cite{IharaMag, SakuraiPD}, implying that the superconductivity has a certain magnetic origin. On the other hand, theoretical studies have pointed out the importance of multi-orbital effects and the possible triplet pairing of Cooper pairs with $p$- and/or $f$-wave symmetry \cite{YanaseMulti}.

However, there are still serious discrepancies in experimental results among some cases even for properties that are critical for understanding superconductivity. One such case of discrepancies is among superconducting phase diagrams. Two kinds of $x$ dependence of $T_{\mbox{c}}$ have been reported \cite{SchaakNature, ChenPRB}, which is probably due to the fact that the Co valence is not determined by the Na content alone; the oxyhydrate includes a significant amount of oxonium ions and the rigorous chemical formula is Na$_{x}$(H$_{3}$O)$_{z}$CoO$_{2}\cdot y'$H$_{2}$O, rather than Na$_{x}$CoO$_{2}\cdot y$H$_{2}$O, which has often been used previously \cite{TakadaJMC}. Thus, we have to consider three compositional parameters, $x$, $z$, and $y'$, although the variation of water content is not significantly large. Milne $et$ $al$. have reported a revised $T$-$s$ ($s$: Co valence) phase diagram with considerion of the oxomium ion \cite{MilnePRL}. Their phase diagram is, however, still far from sufficient because it is not obvious that the properties of the system are governed only by $s$. Indeed, it has been recently elucidated that the ratio of $z/x$ is a crucial parameter governing the phase diagram \cite{SakuraiPD}. Furthermore, their range of Co valence is much lower than those reported by other groups \cite{TakadaJMC, KarppinenCM, TakadaCM, LopezCM}. 

The $H$-$T$ phase diagram ($H$: magnetic field) is also an issue of disagreement because the upper critical fields $H_{c2}$ reported thus far show very large variation. $H_{c2}(T=0)$ perpendicular to the $c$ axis, which was estimated from electrical resistivity measurements, is about 8 T, as determined for samples with $T_{c} = 4.2$ K \cite{ChouPRL} and 4.5 K \cite{SasakiJPSJ}. This value may look reasonable because it is close to the Pauli limit of 1.84$T_{c}$ \cite{ClogstonPRL, ChandrasekharAPL}. However, much higher $H_{c2}$ values have also been reported from magnetic and thermal measurements, such as 15.6 T for a sample with $T_{c} = 3.7$ K \cite{Badica0402235}, 17.1 T for $T_{c} = 4.7$ K \cite{Jin0410517}, 28 T for $T_{c} = 4.3$ K \cite{MaskaPRB}, and 61 T for $T_{c} = 4.6$ K \cite{SakuraiPRB}. These values were obtained by extrapolating the data under lower magnetic fields using the WHH model \cite{WerthamerPR}. However, it is not obvious whether the WHH model and the Pauli limit are adequate for the present system because they are not applicable for spin-triplet superconductivity. Indeed, a theoretical calculation for the compound has predicted spin-triplet superconductivity with $T_{c}$ negligibly reduced by the magnetic field \cite{YanaseMulti}. In other words, the determination of $H_{c2}$ is key to checking the theory.

In the previous study, we drew the phase diagram for the Na$_{x}$(H$_{3}$O)$_{z}$CoO$_{2}\cdot y'$H$_{2}$O system taking $z/x$ as a parameter and keeping the Co valence constant at $s$ = +3.40, $i.e.$, the $s$ = +3.40 cross-section of the three-dimensional $T$-$x$-$z$ phase diagram \cite{SakuraiPD}. Surprisingly, superconductivity appears in separate regions in this section, sandwiching a magnetically ordered nonsuperconducting phase. This phase diagram strongly suggests that magnetic correlation is the origin of superconductivity, but a correlation that is too strong suppresses superconductivity, and instead, magnetic ordering takes its place. In addition, the $H$-$T$ phase diagram for an $s$ = +3.40 sample indicated that the superconducting phase can be transformed to the magnetic phase by applying a high magnetic field. 

Considering the aforementioned progress of research, it is highly desired at present to determine the phase relations for a different Co valence to obtain an overview of the entire three-dimensional phase diagram. Furthermore, high-field magnetic susceptibility measurements for a well-characterized sample are desired to elucidate the phase relationships in the $H$-$T$ diagram. In the present study, we systematically synthesized samples by soft-chemical processes while controlling their composition, and the magnetic susceptibility of one sample was measured up to an extremely high field of 30 T. 

The samples were obtained by a method similar to that previously reported \cite{SakuraiPD}. In the previous study, we added HCl aqueous solution into the system in the final step of the procedure to obtain samples with various $z/x$ values depending on the volume of HCl solution but with the constant $s$ value of +3.40 (HCl series of samples). In the present study, a NaOH aqueous solution was added in the final step of the procedure (NaOH series of samples). First, 1$g$ of Na$_{0.7}$CoO$_{2}$ precursor was immersed in Br$_{2}$/CH$_{3}$CN for 5 days at room temperature, and filtrated samples were immersed in 400 $ml$ of distilled water for 3 days. Then, $v_{\mbox{\small{NaOH}}}$ $ml$ of 0.1 M NaOH aqueous solution was added to the water (instead of $v_{\mbox{\small{NaOH}}}$ = 100 and 500 $ml$, 10 $ml$ and 50 $ml$ of 1 M NaOH solution were really used, respectively), then the samples were kept for 3 more days before performing a final filtration. The samples were characterized and measured after being stored in air with 70\%\ humidity for more than 3 weeks.

\begin{table}
\begin{tabular}{|c|c|c|c|}	\hline
		&	$v$ ($ml$)	&	Na content, $x$	&	Co valence, $s$	\\	\hline\hline
NaOH series	&	0	&	0.332	&	+3.48	\\	\cline{2-4}
		&	10	&	0.346	&	+3.48	\\	\cline{2-4}
		&	20	&	0.354	&	+3.49	\\	\cline{2-4}
		&	30	&	0.357	&	+3.50	\\	\cline{2-4}
		&	40	&	0.358	&	+3.47	\\	\cline{2-4}
		&	100	&	0.368	&	+3.47	\\	\cline{2-4}
		&	500	&	0.380	&	+3.49	\\	\hline\hline
HCl series	&	0	&	0.350	&	+3.41	\\	\cline{2-4}
		&	2	&	0.346	&	+3.40	\\	\cline{2-4}
		&	4	&	0.346	&	+3.41	\\	\cline{2-4}
		&	6	&	0.336	&	+3.40	\\	\cline{2-4}
		&	8	&	0.322	&	+3.40	\\	\cline{2-4}
		&	10	&	0.302	&	+3.36	\\	\hline
\end{tabular}
\caption{Na content and Co valence of the NaOH and HCl series of samples as determined by chemical analyses. The $v$ value represents the volume of added NaOH/HCl aqueous solution.
\label{CA}}
\end{table}

\begin{figure}
\begin{center}
\includegraphics[width=5cm,keepaspectratio]{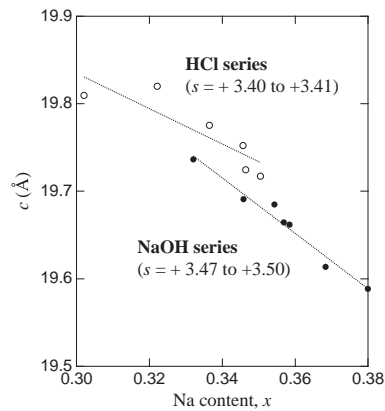}
\end{center}
\caption{
Lattice parameter, $c$, of NaOH (full circles) and HCl (open circles) series of samples. The dotted lines are guides.
\label{LP}}
\end{figure}

The sample with $v_{\mbox{\small{NaOH}}}$ = 500 $ml$ had the largest Na content and a very small amount of the secondary phase of anhydrate Na$_{x}$CoO$_{2}$ was detected in its XRD pattern (the ratio of the peak intensities of the secondary phase and the main phase was about 0.08\%). This suggests that the solid solution terminates near this composition. All other NaOH samples were identified as being of a single phase by XRD analysis. The $c$-axis length decreases with increasing Na content as shown in Fig. \ref{LP} confirming the substitution of the smaller Na$^{+}$ ion for the larger H$_{3}$O$^{+}$ ion.

As discussed previously, the following two side reactions should be considered for the hydration process in addition to the intercalation of the water molecules.
\begin{equation}
\mbox{Na}_{a}\mbox{CoO}_{2} + b\mbox{H}_{2}O \rightarrow \mbox{Na}_{a-b/2}\mbox{(H}_{3}\mbox{O)}_{b/2}\mbox{CoO}_{2} + b/2\mbox{\small{NaOH}}
\label{IonEx}
\end{equation}
\begin{equation}
\mbox{Na}_{a}\mbox{CoO}_{2} + b\mbox{H}_{2}\mbox{O} \rightarrow \mbox{Na}_{a}\mbox{(H}_{3}\mbox{O)}_{2b/3}\mbox{CoO}_{2} + b/6\mbox{O}_{2}
\label{REDOX}
\end{equation}
According to reaction \ref{IonEx}, the sodium ion content (the oxonium ion content) of the final product is expected to increase (decrease) with increasing $v_{\mbox{\small{NaOH}}}$. On the other hand, reaction \ref{REDOX} suggests a decrease in the Co valence with increasing $v_{\mbox{\small{NaOH}}}$ \cite{TakadaJMC}. In Table \ref{CA}, results of the chemical analysis for the NaOH series of samples are shown in comparison with the data for the previous HCl series of samples \cite{SakuraiPD}. As can be seen in the table, the Na content increases systematically as a function of $v_{\mbox{\small{NaOH}}}$ in accordance with reaction \ref{IonEx}.

On the other hand, the analytical data of the Co valence are not fully understandable. First, there is a significant difference in Co valence (and in sodium content though the difference is much smaller) between samples of the two series with $v_{\mbox{\small{HCl}}}$ = 0 $ml$ and $v_{\mbox{\small{NaOH}}}$ = 0 $ml$. These samples were from different batches but were prepared by essentially the same procedure and should have the same analytical values. The difference in Co valence is not negligible and it may suggest that there are still unknown factors in the hydration process and we have not fully controlled the process. Second, the Co valence is kept almost constant at $\sim$+3.48 in the NaOH series of samples independent of $v_{\mbox{\small{NaOH}}}$. Reaction \ref{REDOX} proceeds under an alkaline condition and, thus, a systematic decrease in Co valence is expected with increasing $v_{\mbox{\small{NaOH}}}$. This is not really the case, which implies a more complicated situation exists. In spite of these uncertainties, the analytical data of the NaOH series of samples are quite favorable for our further experiments because from this new series of samples we can obtain a different cross section of the $T$-$x$-$z$ phase diagram with $s$ = $\sim$+3.48.

\begin{figure}
\begin{center}
\includegraphics[width=8cm,keepaspectratio]{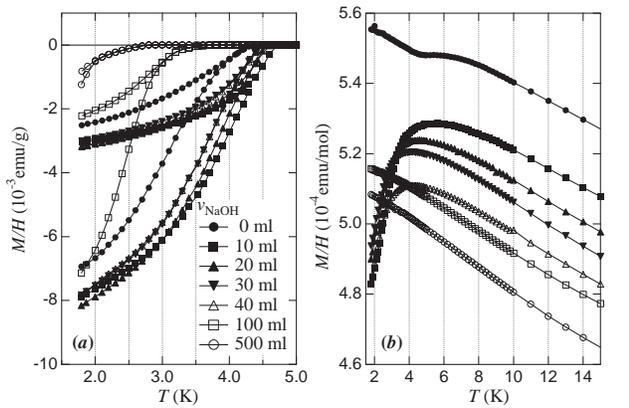}
\end{center}
\caption{
Magnetic susceptibility of the NaOH series measured under 0.001 T (a) and 7 T (b). The data were collected under zero-field cooling and field cooling conditions, although only FC data are shown in the panel of (b).
\label{chi}}
\end{figure}

\begin{figure}
\begin{center}
\includegraphics[width=5cm,keepaspectratio]{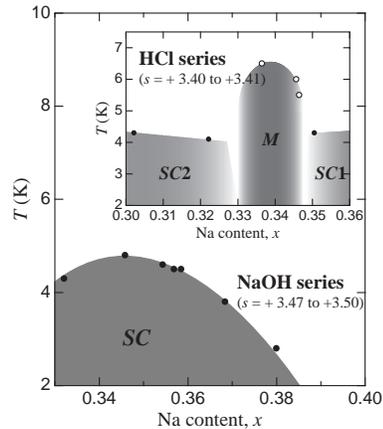}
\end{center}
\caption{
Superconducting phase diagram determined from Fig. \ref{chi}(a), which is a cross section of the three-dimensional $x$-$z$-$T$ phase diagram at $s$ = $\sim$+3.48. The inset shows the cross section at $s$ = $\sim$+3.40 obtained from the HCl series of samples \cite{SakuraiPD}. $SC$ and $M$ denote the superconducting and magnetically ordered phases, respectively.
\label{PD}}
\end{figure}

All of the NaOH series of samples show superconductivity with $T_{c}$ depending on $x$ as seen in Fig. \ref{chi}(a). From these results, the phase diagram is depicted as shown in Fig. \ref{PD}. It should be stressed again that the Co valence is kept almost constant at $\sim$+3.48 and the change in $T_{c}$ is caused by the isovalent substitution between the Na$^{+}$ and H$_{3}$O$^{+}$ ions. The superconductivity appears even for $0.33 \leq x \leq 0.35$, where the magnetic phase is located in the cross section for the HCl series of samples as shown in the inset of Fig. \ref{PD} \cite{SakuraiPD}. This remarkable difference should be caused by the difference in Co valence between the two series of samples. Namely, the region of the $SC$1 phase obtained for the HCl series seems to extend to the lower $x$ direction when the Co valence increases from +3.40 to +3.48. 

The superconductivity of the system depends strongly on the substitution between Na$^{+}$ and H$_{3}$O$^{+}$ (on the $z/x$ ratio) as seen in both cross sections of Fig. \ref{PD} and this suggests that the shape and/or size of the Fermi surface of the present compound is substantially affected by the substitution. Band calculations have led to the proposal of the presence of small hole-like pockets near the K-points as well as a large cylindrical Fermi surface around the $\Gamma$-point \cite{SinghPRB}. The size of the hole-like pockets is expected to be quite sensitive to the thickness of the CoO$_{2}$ layer \cite{MochizukiJPSJ, YadaJPSJ}. With increasing H$_{3}$O$^{+}$ content, the oxygen ions in the CoO$_{2}$ layer will be attracted less strongly to the Na$^{+}$/H$_{3}$O$^{+}$ ions moving toward the Co ions due to the expansion of the distance between the oxygen atoms and the Na$^{+}$/H$_{3}$O$^{+}$ plane. Thus, the substitution of H$_{3}$O$^{+}$ for Na$^{+}$ is expected to make the CoO$_{2}$ layer thinner and the size of the hole-like pockets larger.

Compared with the Na$^{+}$/H$_{3}$O$^{+}$ exchange, the less pronounced influence of the Co valence (carrier density) on superconductivity is worth noting. This also suggests the presence of two kinds of Fermi surface because the carrier density is expected to have a marked effect on the superconductivity if a single Fermi surface is assumed. On the other hand, angle-resolved photoemission spectroscopy (ARPES) experiments have shown a single Fermi surface on anhydrate Na$_{x}$CoO$_{2}$ without the presense of any hole pockets \cite{HasanPRL, YangPRL, Yang0501403, Hasan0501530}. This result is important because many theories proposed thus far \cite{YanaseMulti, Kuroki0508} are based on multiple kinds of Fermi surface: hole pockets around the K-points or small electron pockets at the $\Gamma$-point \cite{Kuroki0508} in addition to the large hexagonal cylindrical Fermi surface centered at the $\Gamma$-point \cite{SinghPRB}. However, we should note here the fact that the thickness of the CoO$_{2}$ layer (1.78 - 1.80 \AA\ \cite{TakadaNature, TakadaJMC}) in the superconducting hydrate is about 10\%\ smaller than that of the anhydrate (1.93 - 1.97 \AA\ depending on the $x$ value ranging from 0.40 - 0.84 \cite{TakadaJMC, BalsysSSI, OnoJSPM, NakatsugawaJSSC}). This difference in layer thickness can cause a significant modification of the band structure as mentioned above. Indeed, a PES study has indicated a qualitative difference in band structure between the hydrate and the anhydrate \cite{ShimojimaPRB}, and moreover, another PES study has supported the existence of two kinds of Fermi surface for the superconductor \cite{KubotaPRB}.

In the previous study, we found that the superconducting phase is transformed into the magnetic phase under a high magnetic field when it is located in the vicinity of the magnetically ordered $M$ phase in the phase diagram \cite{SakuraiPD, SakuraiISS}. The $v_{\mbox{\small{NaOH}}}$ = 0 $ml$ sample with $x = 0.332$ exhibits the transformation in question at $\sim$ 4 K under $\sim$ 4 T, and indeed, its susceptibility has a bend at 6 $\sim$ 7 K under 7 T as seen in Fig. \ref{chi}(b) indicating magnetic ordering. On the other hand, the $v_{\mbox{\small{NaOH}}}$ = 100 and 500 $ml$ samples do not show the anomaly under 7 T which is characteristic of the magnetic transition. Instead, they seem to show signs of superconductivity with saturation of the susceptibility in the low-temperature region, although it is unclear. The present results support the idea that the $SC$1 phase region extends in the lower $x$ direction with increasing Co valence from +3.40 to +3.48. It is not clear at present whether the $M$ phase exists under no or low magnetic field in the $s$ = +3.48 section of the diagram. The $SC$1 and $SC$2 phases are possibly combined with each other under a low magnetic field in the high Co valence section and the $M$ phase may appear only under a high magnetic field in this section.

\begin{figure}
\begin{center}
\includegraphics[width=8cm,keepaspectratio]{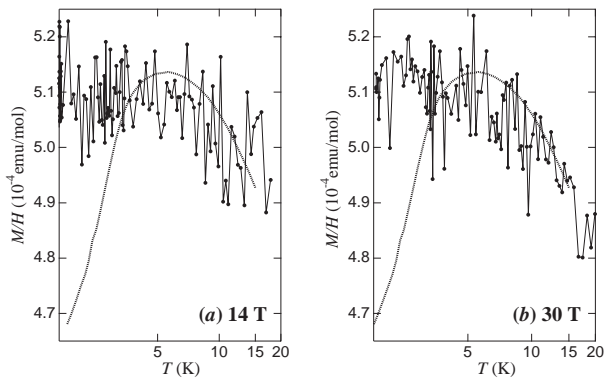}
\end{center}
\caption{
Magnetic susceptibility of the $v_{\mbox{\small{NaOH}}}$ = 10 $ml$ sample measured under 14 T (a) and 30 T (b). The $T$-axis is in the logarithm scale. The dotted lines denote the magnetic susceptibility of the samples under 7 T.
\label{HFMT}}
\end{figure}

 High-field magnetic susceptibilities up to 30 T were measured for the $v_{\mbox{\small{NaOH}}}$ = 10 $ml$ sample having the highest $T_{c}$ among the present samples by an extraction method under the field generated by a hybrid magnet (Gama) at the Tsukuba Magnet Laboratory (the field below 15 T was generated using only a Gama superconducting magnet). As can be seen in Fig. \ref{HFMT}(a), in spite of the rather large dispersion of the data, the susceptibility under 14 T shows a downturn below about 4 K indicating the superconducting transition. It should be noted that this field is much higher than the Pauli limit of 1.84$T_{c}$ = 8.6 T \cite{ClogstonPRL, ChandrasekharAPL}. The superconducting transition was still observed under 20 T and seems to survive even under 30 T (see Fig. \ref{HFMT}(b)) although it becomes less clear due to reducion of the superconducting volume fraction by the field. In addition, as seen in Fig. \ref{chi}(b), $T_{c}$ of the sample under 7 T is much higher than $\sim$2 K or $\sim$2.6 K, which were reported for the samples having $H_{c2}$ = $\sim$8 T \cite{ChouPRL, SasakiJPSJ}. Thus, it is clear that $H_{c2}$ can become higher than the Pauli limit depending on the sample. Our aforementioned results indicate that a compound located close to the magnetic phase in the phase diagram has a low $H_{c2}$ due to the transformation to the magnetic phase under a high magnetic field while one located far from the magnetic phase has a much higher $H_{c2}$. A spin-triplet pairing mechanism with a large $H_{c2}$ value based on strong magnetic fluctuations for holes on the hole pockets \cite{YanaseMulti} has been theoretically proposed, and recently, this theory has been supported experimentally \cite{IharaKnight}. The present finding of $H_{c2}$ much higher than the Pauli limit also supports the validity of this mechanism.

In conclusion, the $x$-$T$ superconducting phase diagram was determined using well-characterized samples and was compared with the previous diagram drawn for $s$ = $\sim$+3.40. The $s$ = +3.48 section of the diagram was different from the $s$ = +3.4 section. It was proved that the superconductivity is very sensitive to the ion exchange between Na$^{+}$ and H$_{3}$O$^{+}$ (the $x/z$ ratio), and less sensitive to the Co valence. This result seems to support the presence of multiple kinds of Fermi surface in which small hole-like pockets near the K points are expected to be changed significantly by the ion exchange. Furthermore, high-field magnetic susceptibility measurements for one sample indicated the survival of superconductivity up to at least 20 T. Such a high $H_{c2}$ supports the recently proposed spin-triplet pairing mechanism both theoretically and experimentally.

%\begin{acknowledgments}
We would like to thank S. Takenouchi and K. Kosuda (NIMS) for the composition analyses, and Y. Yanase (University of Tokyo) for fruitful discussion. This work is supported partially by CREST, JST, and by Grants-in-Aid for Scientific Research from JPSP and MEXT (1634011, 16076209).
%\end{acknowledgments}

% If in two-column mode, this environment will change to single-column
% format so that long equations can be displayed. Use
%\begin{widetext}
%\end{widetext}

% Surround figure environment with turnpage environment for landscape
% figure
% \begin{figure}
% \end{figure}

%\pagebreak

\end{document}